%% file: industry.tex
\documentclass[10pt,conference]{IEEEtran}
\usepackage{url}
\usepackage{cite}
\usepackage{amsmath,amssymb,amsfonts,booktabs,array}
\usepackage{algorithmic}
\usepackage{graphicx}
\usepackage{textcomp}
\usepackage{xcolor}
\usepackage{listings}
\usepackage{mdframed}

\usepackage{marvosym}
\usepackage[symbol]{footmisc}

\include{mine}

\def \TOOLNAME {\textsc{Ratel}}
\begin{document}

\title{Industry Practice of Coverage-Guided Enterprise-Level DBMS Fuzzing}
\author{Mingzhe Wang$^{1}$, Zhiyong Wu$^{1}$, Xinyi Xu$^{1}$, Jie Liang$^{1}$, Chijin Zhou$^{1}$, Huafeng Zhang$^{2}$, Yu Jiang$^{1}\textsuperscript{\Letter}$\\
KLISS, BNRist, School of Software, Tsinghua Universiy$^{1}$\\Huawei Technologies Co. Ltd, Beijing, China$^{2}$}
\maketitle

\footnotetext{\Letter \  Yu Jiang is the correspondence author.}

\begin{abstract}
As an infrastructure for data persistence and analysis, Database Management Systems (DBMSs) are the cornerstones of modern enterprise software. To improve their correctness, the industry has been applying blackbox fuzzing for decades.
Recently, the research community achieved impressive fuzzing gains using coverage guidance. However, due to the complexity and distributed nature of enterprise-level DBMSs, seldom are these researches applied to the industry.

In this paper, we apply coverage-guided fuzzing to enterprise-level DBMSs from Huawei and Bloomberg LP. In our practice of testing GaussDB and Comdb2, we found major challenges in all three testing stages. The challenges are collecting precise coverage, optimizing fuzzing performance, and analyzing root causes. In search of a general method to overcome these challenges, we propose \TOOLNAME{}, a coverage-guided fuzzer for enterprise-level DBMSs. With its industry-oriented design, \TOOLNAME{} improves the feedback precision, enhances the robustness of input generation, and performs an on-line investigation on the root cause of bugs. As a result, \TOOLNAME{} outperformed other fuzzers in terms of coverage and bugs. Compared to industrial black box fuzzers SQLsmith and SQLancer, as well as coverage-guided academic fuzzer Squirrel, \TOOLNAME{} covered 38.38\%, 106.14\%, 583.05\% more basic blocks than the best results of other three fuzzers in GaussDB, PostgreSQL, and Comdb2, respectively. More importantly, \TOOLNAME{} has discovered 32, 42, and 5 unknown bugs in GaussDB, Comdb2, and PostgreSQL. 
%All the bugs in GaussDB are confirmed and fixed by the maintainers.
\end{abstract}


\section{Introduction}
As an infrastructure for data persistence and analysis, Database Management Systems (DBMSs) are widely applied in the modern software stack, especially enterprise software. In recent years, their reliability and security have gained traction in academic research as well as in industry. 

To improve the correctness of DBMSs, the industry has been applying blackbox fuzzing for decades. Basically, to perform fuzz testing, a fuzzer generates a number of random inputs and sends them to the target system for execution. If anomalies such as crashes, timeouts, internal errors, or incorrect results are detected, then the triggering input is saved. Developers can use the input to reproduce the anomaly, investigate the root cause, and fix the bug. For example, SQLsmith~\cite{SQLsmith} triggers system bugs by continuously generating random SQL queries; RAGS~\cite{DBLP:conf/vldb/Slutz98} detects logic bugs by comparing the results of a query on multiple DBMSs; SQLancer~\cite{SQLancer} detects logic bugs by constructing an invariant oracle from different angles~\cite{NoREC,PQS}.
The effectiveness of blackbox fuzzing has been proven by many previously-unknown bugs from the industry's practice: more than 100 bugs were found by SQLsmith, and more than 400 bugs were found by SQLancer.

Recently, coverage-guided fuzzing has become a hot topic in the research community, and impressive gains over the conventional blackbox fuzzing methods are observed \cite{PAFL, zeror}. Coverage-guided fuzzing introduces feedback into the fuzzing loop: more than blindly generating random inputs, a coverage-guided fuzzer collects the execution trace of an input to recognize interesting inputs triggering new program behaviors; the interesting ones are preserved for further mutations, while the mundane ones are discarded. For example, Squirrel~\cite{SQUIRREL} combines coverage-based fuzzing and model-based generation. It performs type-based mutations on intermediate representations and optimizes for semantic correctness with additional analysis. Coverage-guided DBMS testing is especially effective in libraries. For instance, when fuzzing SQLite, Squirrel achieved 7.7x edge coverage gains over the typical blackbox fuzzer SQLsmith. 

However, the performance of coverage-guided DBMS fuzzing on libraries does not match with that of enterprise-level DBMSs. For example, Squirrel discovered 51 new bugs on SQLite, but it failed to discover any bugs on PostgreSQL. This is counter-intuitive: SQLite is released as one source file and has around 750 bugs \cite{SQLite-Bugs}; while for PostgreSQL, a distributed system with millions of lines of code, its bug ID has already exceeded 16,500 \cite{PostgreSQL-Bugs}. Different from library-level DBMSs, enterprise-level DBMSs are much more difficult to test due to extra complexity and distributed nature. For example, GaussDB, Huawei's proprietary DBMS, extends PostgreSQL with enterprise-level features such as advanced high-availability and geo-based sharing. This is implemented with six distributed components in over 10 million lines of code. Therefore, no bugs were discovered when we primitively applied fuzzing to GaussDB.

In our practice, the gap can be divided into three aspects.
First, the coverage collected by conventional fuzzers is low in quality. Commonly, enterprise-level DBMSs execute a diversity of components, and each component contains a number of basic blocks. Conventional fuzzers experience severe collisions in feedback, or cannot collect coverage from multiple components at all.
Next, the SQL generated by existing fuzzers cannot be accepted by enterprise-level DBMSs. Generative fuzzers can produce SQL statements, but the generated inputs are neither correct nor complete, given the diverse syntax of enterprise-level DBMSs.
Even if a fuzzer can trigger a bug, the root cause analysis is still a tough task --- enterprise-level DBMSs' contextual dependencies are difficult to model, especially given a sequence of random inputs generated with non-deterministic mutation algorithms.

In this paper, we first present the challenges we faced in our practice, and then propose general solutions for them. After resolving the challenges, we implement \TOOLNAME{}, a coverage-guided fuzzer targeted at real enterprise-level DBMSs in the industry. \TOOLNAME{} tackles three major challenges in existing DBMS fuzzing techniques: 1) improve the feedback precision on large-scale distributed systems with inter-binary coverage linkage and bijective block mapping; 2) enhance SQL generation with robustness-oriented strategy; 3) investigate the root cause of anomalies with on-line analysis and deduplication.

We implemented \TOOLNAME{} in Rust, C, and C++. The main fuzzer is implemented in Rust using the tokio asynchronous framework and the input generation by Rui et al. \cite{SQUIRREL}. For precise instrumentation, we also implemented the compiler part based on LLVM \cite{LLVM} with additional passes and pipeline tweaks.
We use \TOOLNAME{} to test GaussDB continuously and discovered 32 previously-unknown bugs in GaussDB. 
%all of which were confirmed and fixed by the maintainers.
For proper evaluation, we further extended our evaluation with Bloomberg LP's distributed DBMS Comdb2 and the original PostgreSQL where GaussDB derived. We also discovered 42 bugs of Comdb2 and 5 bugs of PostgreSQL from their codebases.

To summarize, our main contributions are as follows:
\begin{itemize}
    \item We adapt several fuzzers to enterprise-level DBMSs and identify three challenges encountered in our practice to fuzz DBMSs. The challenges are 
        imprecise coverage collection on distributed systems,
        fragile input generation of complex SQL dialects,
        and unreproducible bug reports on stateful systems.
    \item We propose \TOOLNAME{} to solve the challenges as a general solution for coverage-guided fuzzing on enterprise-level DBMSs. We improve the feedback precision, enhance the robustness of input generation, and perform on-line investigation on the root cause of bugs.
    \item The final results show that \TOOLNAME{} effectively improved existing DBMS fuzzing works on enterprise-level DBMSs. 
    Compared to SQLsmith, SQLancer, and Squirrel, it covered 38.38\%, 106.14\%, 583.05\% more basic blocks than the best results of other three fuzzers in GaussDB, PostgreSQL, and Comdb2, respectively. More importantly, 
    79 unknown bugs are discovered by it.
\end{itemize}

% The rest of this paper is organized as follows. Section~\ref{background} introduces the background of DBMS and DBMS fuzzing, and details of three fuzzers chosen in this paper for industry practice. Section~\ref{fuzz_procedure} illustrates the general procedure of fuzzing databases. Section~\ref{challenges_solutions} elaborates the challenges we encountered and our solutions for them. 
% Section~\ref{challenges_solutions}

\section{Background}
\label{background}
In this section, we first give a brief introduction to DBMS and give an example to illustrate the complexity of enterprise-level DBMSs.
Then, we introduce some existing solutions of DBMS fuzzing.
Finally, we detail three state-of-the-art fuzzers, i.e. SQLsmith, SQLancer, and Squirrel, which are chosen for our industry practice.

\subsection{Database Management System (DBMS)}
\begin{figure}[htbp]
    \center
    \includegraphics[width=0.8\linewidth]{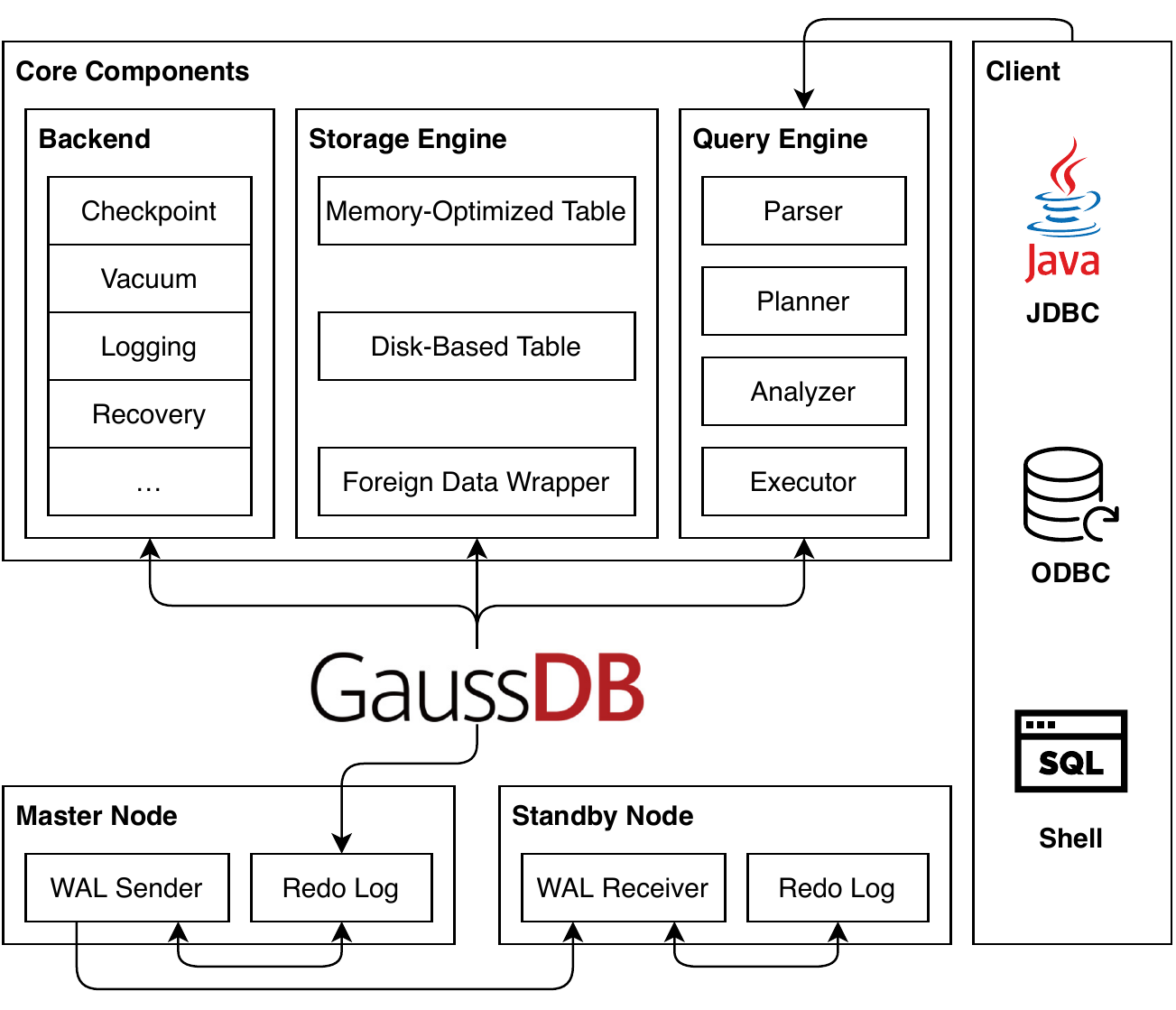}
    % \caption{Classic setup of PostgreSQL. The \textit{primary} host handles all writes, and replicates all modifications to the \textit{standby} host. The \textit{client} host sends queries to both type of hosts. }
    \caption{Interactions between different processes/threads in GaussDB. }
    \label{fig:gauss-workflow}
\end{figure}

Database management system (DBMS) is the software that interacts with end users and applications. Different from libraries and utilities, most DBMSs are large and complex distributed systems. They generally contain hundreds of components, within which complex interactions reside during runtime. 
Fig.~\ref{fig:gauss-workflow} gives a high-level overview of interactions on a classic database cluster setup of GaussDB~\cite{gaussdb-doc,DBLP:journals/pvldb/AvniAAABGGLLLMM20}. The \textit{server} listens on a socket to accept queries from \textit{clients}. Upon receiving queries, it parses them by \textit{Query Engine}, and searches results from \textit{Storage Engine}. For replication, \textit{Master Node} sends replicated modifications to \textit{Standby Nodes}. Master node starts \texttt{WALsender} to perform replication, and multiple standby nodes start \texttt{WALreceiver} to receive write-ahead logs (WAL) from \text{Master Node} and maintain their \textit{RedoLog} states. Meanwhile, the \texttt{backend} process is spawned periodically for certain purposes, e.g. \textit{Vacuum} for garbage analysis and collection, \textit{Checkpoint} for transaction log persistence. 
From the interactions of GaussDB, it can be seen that there are many kind of processes interweaving with main process in enterprise-level DBMSs, some of which are even deployed on different hosts for high-availability and scalability.

\subsection{Fuzzing DBMSs}

Due to the complexity and distributed nature of DBMS, ensuring its reliability and security is difficult.
Therefore, many DBMS fuzzers have sprung up in the past years and tried to find bugs in different aspects. Generally, two approaches were taken by them, i.e. blackbox fuzzing and coverage-guided fuzzing.

Blackbox DBMS fuzzing focuses on rapidly constructing a large amount of valid and effective data to pass to the DBMS without the knowledge of the DBMS's source code. For instance, SQLsmith~\cite{SQLsmith} continuously generates syntactically-correct SQL statements and passes them to the DBMS, meanwhile detects whether the DBMS triggers crashes.
In addition, SQLancer~\cite{SQLancer} integrates three different strategies~\cite{PQS,NoREC,TLP} to construct invirant oracle to detect logic bugs. For example, its pivoted query synthesis strategy~\cite{PQS} generates queries of which corresponding result table is supposed to include a specific row, and if the DBMS fails to fetch the row, a logic bug is discovered. 
Additionally, some research works~\cite{DBLP:conf/vldb/Slutz98,jung:apollo} leverage differential testing to detect abnormal behavior in DBMSs. RAGS~\cite{DBLP:conf/vldb/Slutz98} generates and executes queries in multiple DBMSs, meanwhile compares the results of a query on those DBMSs; similar to RAGS, APPOLO~\cite{jung:apollo} generates queries guided by performance difference and compares the results of a query on different versions of the same DBMS. 
All these blackbox DBMS fuzzers have shown good performance on bug discovery in the industry's practice. For example, SQLsmith has found 118 bugs in the past five years\cite{SQLsmith_score_list}, SQLancer has found over 400 bugs in the past two years~\cite{sqlancer_score_list}.

While blackbox fuzzing for DBMSs has gained popularity, coverage-guided fuzzing is hardly applied in the industry. To increase code coverage, coverage-guided fuzzing leverages instrumentation or even program analysis techniques; sometimes, directed fuzzing can also be used to reach certain critical program locations.
Traditional coverage-guided fuzzers, like AFL~\cite{afl}, libFuzzer~\cite{libFuzzer} and Honggfuzz~\cite{honggfuzz}, can be easily applied to DBMS libraries such as SQLite. These fuzzers use mutation-based input generation techniques, and the generated inputs are very unlikely to be syntactically correct. To generate inputs without being blocked by shallow syntax checks, users usually use a customized SQL dictionary. During the mutation stage, fuzzers will randomly replace a part of the generated input with the given SQL keywords. While symbolic execution \cite{DBLP:conf/icse/SAFL, DBLP:conf/ndss/Driller} and enhanced mutation \cite{DBLP:conf/kbse/FairFuzz, InteFuzz} improved the quality of generated inputs, coverage-guided fuzzing are still blocked by shallow checks in most cases.
Recently, Squirrel~\cite{SQUIRREL} has been proposed to combine model-based generation and coverage-guided fuzzing for DBMS. It is built on top of AFL just like other fuzzers. The different part is the use of Bison and Flex for parsing the original SQL statements provided by the user into intermediate representations. By using intermediate representation instead of a plain buffer of bytes, the structural information of the SQL statements are preserved, and further syntactically correct mutations can be performed. Over 60 bugs were discovered by Squirrel.

\subsection{Fuzzers Chosen by This Paper}
\label{sec:fuzzer-chosen}
To improve the correctness of GaussDB, we try to perform fuzzing on typical fuzzers, including both the blackbox and coverage-guided ones. As shown above, many DBMS fuzzers were proposed, and their performance seems to be attractive. However, after analyzing lots of trails and errors, we found that only a few of them can be applied practically for bug hunting. For example, RAGS is not open-source and APPOLO requires a certain version of DBMS and the adaption cost is great.
After considering the investments and the possible returns, we chose three fuzzers as our target in our experiment. Table ~\ref{tab:chosen-fuzzer} shows the characteristics of these fuzzers. We summarize their features as follow:
\begin{itemize}
    \item \textit{SQLsmith} is a random SQL query generator. It detects bugs by checking whether the connection to server is closed. It is easy to adapt to a new DBMS because it only needs to connect to target SQL server.
    \item \textit{SQLancer} is a fuzzer for hunting logic bugs in DBMS. It detects logic bugs by constructing invariant oracle~\cite{PQS,TLP,NoREC} and checking whether results violate semantic logic. Adapting it to a new DBMS may take some time because new logic assertion and syntax generation should be implemented.
    \item \textit{Squirrel} is a fuzzer aiming at finding memory corruption issues in DBMS. It detects memory corruption in a way similar to SQLsmith, i.e. checking whether the connection to server is closed. Adapting it to a new DBMS may take many manual efforts because it requires customizing AST parser, writing grammar rules, and implementing client logic under its framework.
\end{itemize}

\begin{table}[th]
    \small
    \caption{\label{tab:chosen-fuzzer} Features of Chosen Fuzzers}
    \centering
    % \resizebox{1\textwidth}{7mm} {
    \begin{tabular}{llll}
      \toprule
      Fuzzer              &SQLsmith  &SQLancer  &Squirrel   \\
      \midrule
      Syntax Validity     &${\surd}$ &${\surd}$ &${\surd}$  \\
      Semantic Validity   &$\times$  &${\surd}$ &${\surd}$  \\
      Logic Check         &$\times$  &${\surd}$ &$\times$   \\
      Coverage-Guided     &$\times$  &$\times$  &${\surd}$  \\
      Adaption Difficulty &Easy      &Medium    &Hard       \\
      \bottomrule
    \end{tabular}
    % }
\end{table}

\section{How to Fuzz Database}
\label{fuzz_procedure}
\begin{figure}[htbp]
    \center
    \includegraphics[width=0.8\linewidth]{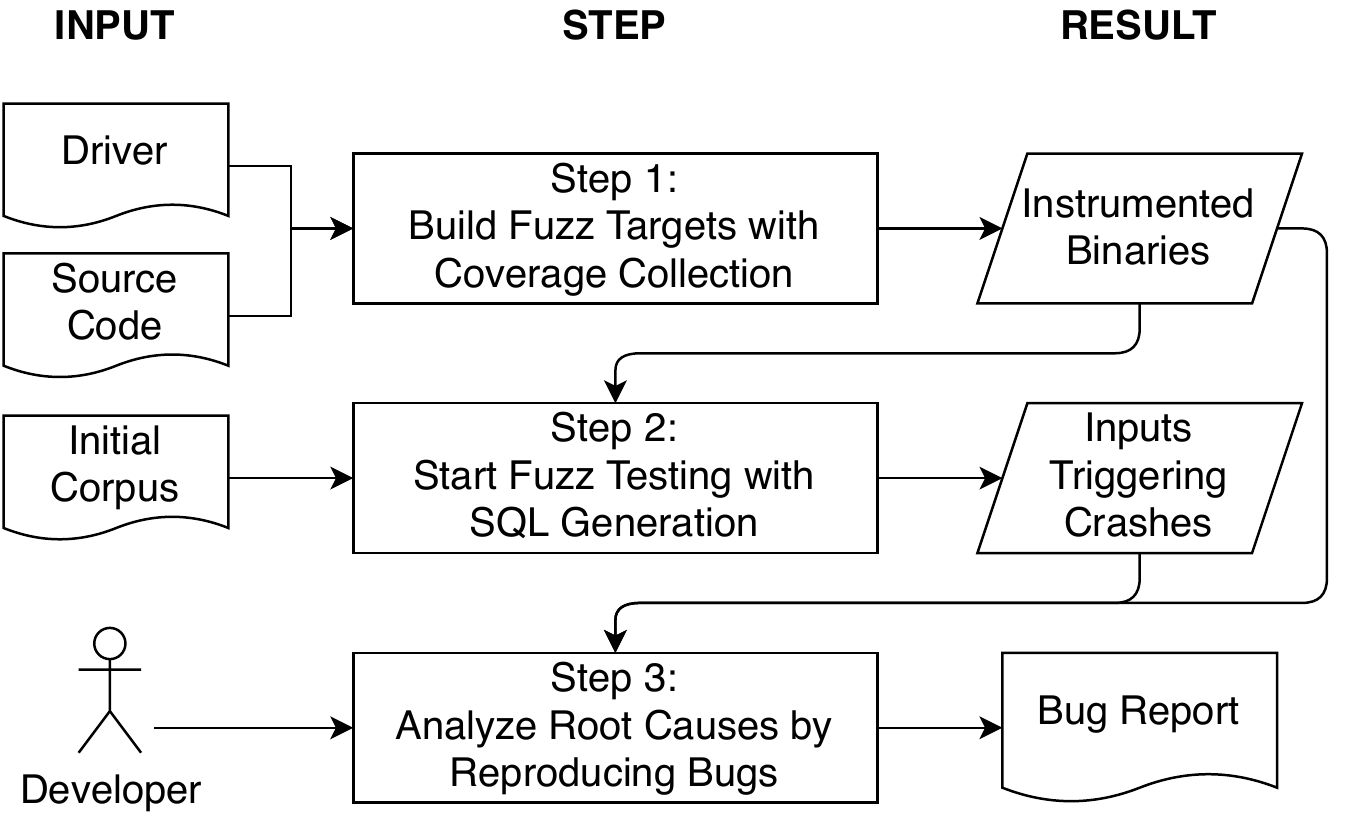}
    \caption{Steps of enterprise-level DBMS testing with coverage guidance.}
    \label{fig:workflow}
\end{figure}

Different DBMSs have different properties, but the procedures to fuzz them with coverage guidance are roughly similar. The steps can be summarized as Fig. \ref{fig:workflow}. First, to collect coverage on executions, the fuzz target should be built into instrumented binaries. Next, to generate correct SQL statements, the high-quality initial corpus is collected for further mutations. During the process, crash-triggering inputs are saved for further inspection. Finally, to analyze the root causes, the abnormal inputs are resubmitted to the binaries to reproduce the bug and create bug reports manually.

\paragraph{Building Fuzz Targets}
As a dynamic testing method, fuzzing requires running the target system with proper instrumentation. Therefore, the first step of fuzzing is building the target project with the source code and the driver.

A test driver is the entrance of fuzz testing, which feeds inputs to the target system. Fuzzers such as SQLsmith and Squirrel embed the query logic into themselves. Fuzzers aiming at improving the generality can use the common interface to send queries, such as JDBC and ODBC.

In addition to building the driver, we also need to compile the source code to generate binary files.
For blackbox fuzzers, general-purpose compilers such as gcc or clang can be used. We enable optimizations for better performance and enable sanitizers for enhanced bug-detection ability. %Address sanitizer is especially important, because it instruments the program to detect memory-safety bugs which can lead to severe security incidents such as arbitrary code execution. Although memory-safety bugs sometimes can be detected by the operating system, the latest ones such as double-free can be difficult to debug. The additional context information recorded by sanitizer, e.g. the stack where the memory is allocated or freed, can be a valuable source to pinpoint the root cause.
For coverage-guided fuzzers, they are required to collect feedback from program executions. One popular method is the compile-time instrumentation. For example, when building the target program for Squirrel's use, we run configuration scripts with \texttt{CC} and \texttt{CXX} pointed to \texttt{afl-gcc} and \texttt{afl-clang++}. These tools hijack the execution of the default compiler, and inserts a constant-sized counter region to log the program's execution trace. On the entrance of each basic block, an additional logging code is executed, which increments the corresponding counter according to the transition of basic blocks.

\paragraph{Generate SQL Queries}
Automatically-generated SQL queries can cover complex and corner cases, which reveal bugs beyond simple unit tests or hand-written tests. There are two mainstream methods to generate SQL queries automatically.
1) Buffer-based mutation, where SQL statements are treated as a plain buffer of bytes. This method randomly changes components of the byte buffer with substitution, deletion, and insertion; sometimes, multiple buffers are spliced together. The buffer-based mutation is relatively simple to implement and can automatically generate a large number of SQL, but most of the generated SQL statements are syntactically invalid.
2) Syntax-based mutation. Instead of simply mutating on the byte level, this method
constructs SQL statements by building abstract syntax trees (AST). This method relies on corpora of partially constructed SQL statements in the form of ASTs, covering different types of SQL syntax. To generate SQL statements, the original SQL statements are first parsed into ASTs. Next, transformations such as splicing and deletion are performed on the nodes of the ASTs. Finally, the transformed ASTs are serialized into SQL statements. Although this type of generation cannot ensure the correctness of semantics, the generated inputs are syntactically-correct at least. Consequently, early rejections at parsers are avoided, and deeper logic in the target system can be exercised. Many tools are implemented following this method, including SQLsmith, SQLancer, and Squirrel.

    % \paragraph{Start fuzzing and monitor the process}
    % When we can input SQL statements into DBMS continuously and execute them, our next main task is to start fuzzing and monitor the process under test. The whole process can be divided into two steps. First, fuzzer can mutate the initial seed and generate a large number of SQL statements. Different Fuzzers have their own different SQL statement generation methods. The specific comparison is shown in the \texttt{tablex}.
% \begin{table}[h]

% \renewcommand\arraystretch{1.4} 
% \normalsize
% \begin{tabular}{c|l|l|l}
% \hline
% \textbf{Fuzzer} &
%   \multicolumn{1}{c|}{\textbf{Types of SQL}} &
%   \multicolumn{1}{c|}{\textbf{Semantic }} &
%   \multicolumn{1}{c}{\textbf{Grammar}} \\ \hline
% SQLsmith &  &  &  \\
% SQLancer &
%   \begin{tabular}[c]{@{}l@{}}\end{tabular} &
%   &
%   \\
% Squirrel &  &  &  \\ \hline
% \end{tabular}

% \end{table}
    % After generating a large number of SQL statements, it is necessary to input the SQL statements into DBMS for execution with the driver written before. During the execution of DBMS, fuzzer need to monitor the execution process of DBMS. According to different monitoring levels, it can be divided into Class, 1. Only monitor whether the execution results are normal, and do not monitor the components involved in the SQL execution process, such as SQLancer; 2. Monitor the execution process of SQL in DBMS, but only monitor whether the function is abnormal? For example, Squirrel 3. Monitor the communication threads between different components involved in the execution of SQL in DBMS?\TODO
    %这里描述问题很大，不知道怎样才能正确的区分
    
\paragraph{Analyze Root Causes}
When executing randomly-generated SQL statements, unexpected behaviors indicating bugs may occur. To analyze root causes and fix the bug, A fuzzer must contain a test oracle to detect execution anomalies, or the bug-triggering input may be discarded. Additionally, the related context must be collected too, or the bug will be hard to reproduce and analyze.

For fuzzers targeted at different types of bugs, different types of anomalies may be monitored, and the context information to record also differs. According to the way of finding bugs and monitoring anomalies, we can roughly divide the fuzzing tools into two categories. To detect semantic bugs, tools such as SQLancer implement a test oracle by verifying the returned values. If the results are unexpected, the original query and description of mismatched semantics are logged for further inspection. To detect implementation bugs, tools such as Squirrel looks for crashes and then record the input triggering them.

Although these tools are targeted at different types of bugs, they all monitor for execution anomalies which provide valuable context for pinpointing the culprits. After discovering the anomalies, developers should analyze and classify the anomalies to find root causes. Sometimes, the bug can be reproduced with the anomaly-triggering input for a better understanding of the situation.

\section{Challenges And Solutions}
\label{challenges_solutions}

\begin{figure*}[tbp]
    \center
    \includegraphics[width=\linewidth]{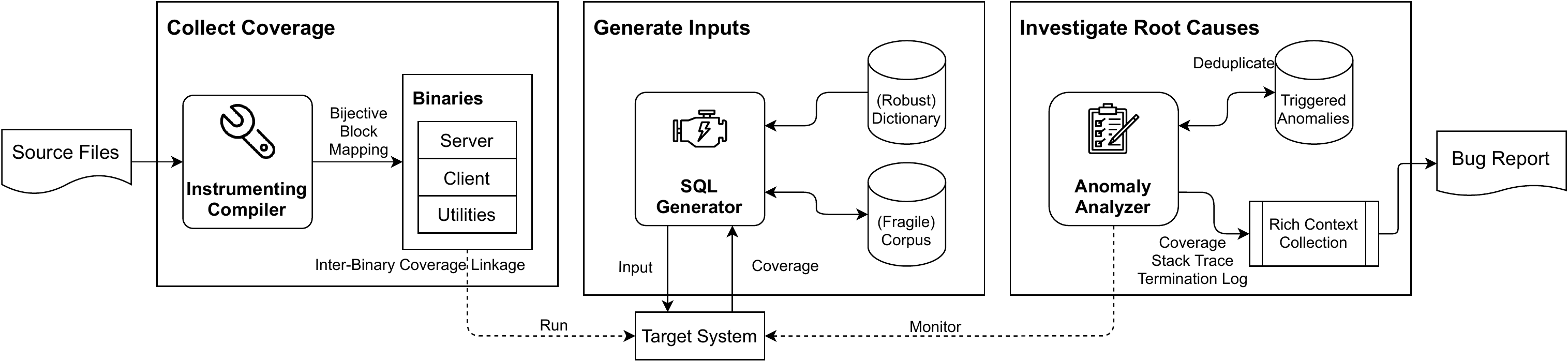}
    \caption{Overview of challenges and solutions in each step. 1) When building fuzz targets, the precision of coverage collection can be improved with bijective block mapping for accuracy and inter-binary coverage linkage for completeness. 2) When generating SQL queries, the fragile SQL corpus-based mutation can be strengthened with dictionaries. 3) When investigating root causes, the analysis of unreproducible anomalies can be performed on-line with deduplication and rich context collection.}
    \label{fig:challenge-solution}
\end{figure*}

As Section~\ref{sec:fuzzer-chosen} lists, we chose the most popular DBMS testing tools, i.e. SQLsmith, SQLancer, and Squirrel to test enterprise-level DBMSs. Despite their wide usage in real-world projects, we still faced considerable challenges when using the coverage-guided fuzzer on enterprise-level DBMSs. As summarized in Fig. \ref{fig:challenge-solution}, in the following, we present the details of challenges as well as the solutions in each step.

\subsection{Imprecise Coverage Collection}
Compared to well-tested library-based DBMSs, the introduction of enterprise-level features in industrial DBMSs leads to extra logic, and inevitably, more components. For example, each node of GaussDB can support at most 8,000 concurrent connections. To support business on large scale, GaussDB implements clustered data nodes for storage and computation, and two to four additional controller nodes are introduced to coordinate them. This simplified version is implemented with a code of 466K basic blocks, not to mention deployments real-life deployments with components for high-availability, monitoring, and management.

Such a complex system poses challenges for fuzzers in terms of efficiency. While blackbox fuzzers are robust against such a complex system, their efficiency is quite limited. When running blackbox fuzzers on DBMSs, we find new code coverage takes many executions to discover. Take SQLsmith as an example, although the number of covered number of basic blocks grew continuously, the growth is extraordinarily slow compared to coverage-guided methods. We observed similar results on SQLancer.

Coverage-guided fuzzer Squirrel improves efficiency by discovering the interesting inputs and assigning more power to mutate them. Compared to blackbox fuzzers, the efficiency of Squirrel is greatly improved, but only for a short time: the coverage of Squirrel quickly reached a plateau. By collecting the number of covered basic blocks over time, we found that a bottleneck is reached soon because its newly generated inputs hardly improve the coverage.

To solve the problem of efficiency and bottleneck, we investigated the causes behind it. The problem can be mainly attributed to the loss of precise coverage feedback collection.
For example, SQLsmith and SQLancer test DBMS with blackbox methods. In other words, though they can determine whether executions of SQL statements are normal, they do not have any coverage feedback. As a result, despite being able to generate syntactically correct SQL statements, these fuzzers could only expand the coverage blindly. The SQL statements generated by the fuzzers are in a narrow-range, which leads to difficulties in covering new basic blocks or finding bugs in deep logic.

Squirrel introduces sparse feedback in AFL to guide fuzzing on DBMSs. It hashes the branch (transitions of basic blocks) to one position in a 256KB bitmap. However, the coverage feedback of Squirrel is incomplete and inaccurate. 
Enterprise-level DBMSs typically contain multiple processes running on multiple hosts, while the instrument mechanism of AFL is designed for testing a single process. As a result, Squirrel can only collect the coverage of the main process of the DBMS server, but for other components such as the logic of the client, it simply discards tracking them. Even if all the components are tracked, AFL's branch coverage itself is imperfect due to the hash collision issue. Despite Squirrel has expanded AFL's bitmap from 64KB to 256KB, for GaussDB which has about 466K basic blocks, the number of branches is much larger than the size of the bitmap. Consequently, the hash collision is still serious: for the best possible hash assignment scheme, 1.82 basic blocks still share the same counter. The edge-based counter-mapping also worsens the conflicts compared to basic-block-based mapping. We can expand the bitmap at the cost of a greater overhead of scanning the bitmap, yet the reduced collision is minuscule.  %\cite{collafl}.

As a result, in order to adapt feedback fuzzing to DBMSs effectively, recording the code coverage accurately and completely has become one of the main challenges.

\begin{figure}[htbp]
    \centering
    \includegraphics[width=0.8\linewidth]{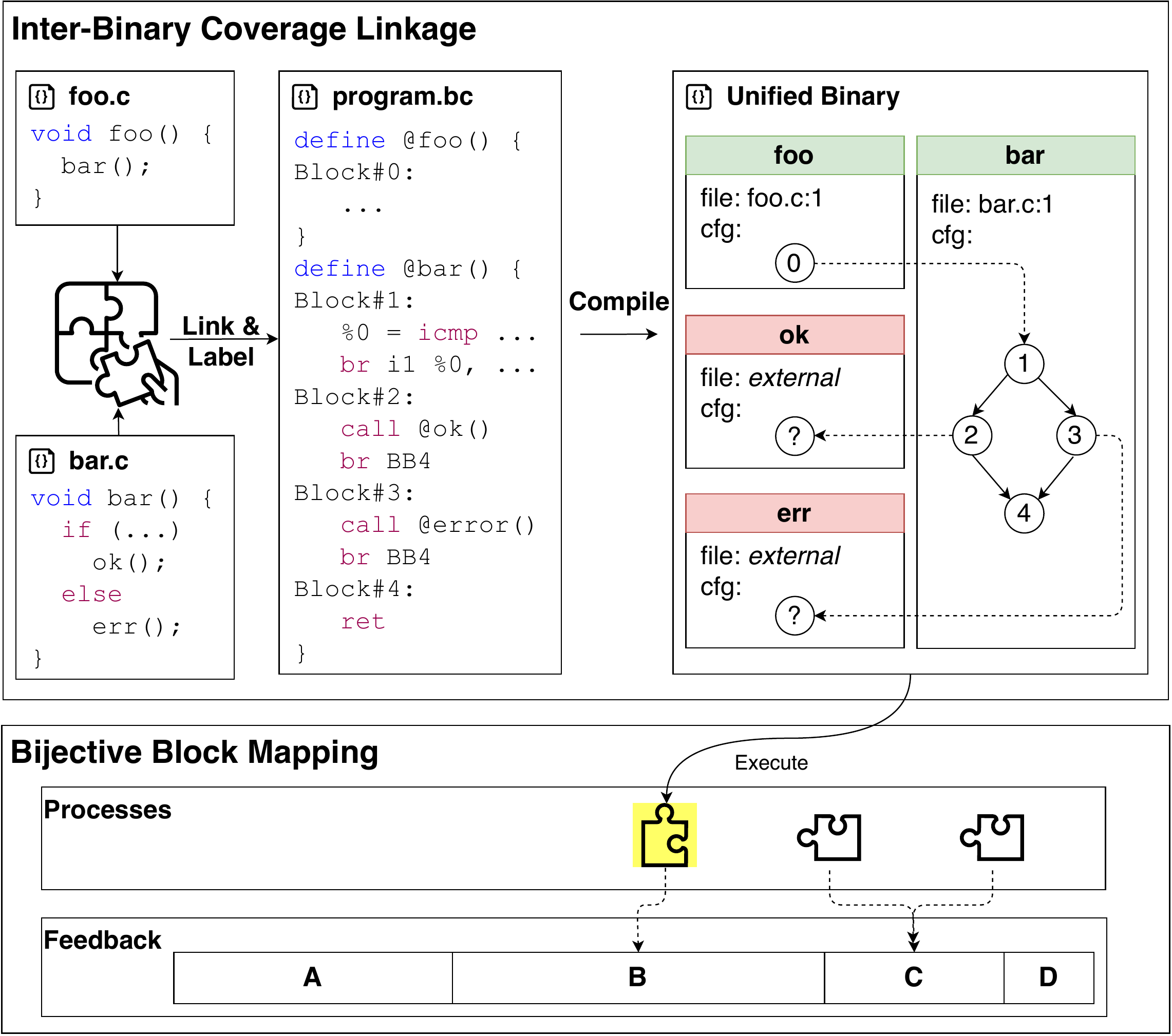}
    \caption{Precise coverage collection with inter-binary coverage linkage and bijective block mapping.}
    \label{fig:instru}
\end{figure}

\begin{mdframed}
\textit{Solution:}

\textit{Collect precise feedback by inter-binary coverage linkage and bijective block mapping.}
\end{mdframed}

Bijective block mapping ensures accuracy. With a global view of the target DBMS, the detailed control flow graphs and call graphs maps each basic block to a unique counter. The counters are placed in the feedback in order, and the hash collision is consequently eliminated. Because of the one-to-one matching between the counters and the basic blocks, we do not introduce the overhead of extra counters like hash-based mapping schemes. To enrich the block-based coverage, we identify the critical branch, where the branch's source has multiple successors and the branch's destination has multiple predecessors, and the critical branch is replaced with a dummy basic block. This design enhances the accuracy of block-based coverage without incurring extra runtime penalties.

Inter-binary coverage linkage ensures completeness. As Fig. \ref{fig:instru} illustrates, we identify basic block information of each binary in the target DBMS in a global view. Like linkers which layout object files in the compilation, inter-binary coverage linkage collects the list of target binaries, fetches the basic block information computed at compile-time, and generates a conflict-free layout to store the coverage among a dynamic set of processes. Therefore, we guarantee the completeness by monitoring each active binary and accounting the coverage for each binary. Consequently, when determining whether an input is interesting, \TOOLNAME{} will not discard interesting inputs by mistake.

\subsection{Fragile Input Generation}
Most DBMS fuzzers leverage handcrafted AST as an input model to generate a greater proportion of valid inputs. Especially, Squirrel uses not only handcrafted AST but also customized semantic rules to ensure semantic validity. These generation-based methods have shown significant effectiveness in DBMS bug discovery~\cite{sqlancer_score_list, SQUIRREL}. 

However, even with enormous manual efforts in modeling a new DBMS's syntax, the model is still fragile due to the complexity of DBMS. DBMSs generally conform to ISO SQL standard~\cite{iso_standard}, which contains 16 parts with thousands of pages and is still in evolution, therefore, constructing an input model is labor-consuming. Take PostgreSQL as an example: SQLsmith only targets PostgreSQL's \texttt{SELECT} statement, and its grammar model already has 42 elements; SQLancer use over 8,000 lines of Java code to generate syntactically-correct test cases; Squirrel uses over 33,000 lines of C++ code to translate between SQL statement and AST.

Another reason is the diversity of different DBMSs. In addition to basic database-related functionality, DBMSs often have their own specific features. For example, despite being a derived product from PostgreSQL, GaussDB still has vastly different SQL dialects. As Listing~\ref{lst:diff-code} shows, in terms of \texttt{CREATE TABLE}, GaussDB adds a special field for choosing its adaptive compression algorithms while shutting off functionalities of inheriting other tables and using extra methods. As a result, its proprietary input model is difficult to transfer to other DBMS.

\setlength{\fboxsep}{1pt}
\begin{lstlisting}[
    caption = {[upper/lower text]%
               \begin{tabular}[t]{@{}l@{}}
                 Different SQL dialect when creating a table \\
                 (PostgreSQL in yellow, GaussDB in Red) \\[.5\normalbaselineskip]
               \end{tabular}},
    label=lst:diff-code]

CREATE [ [ ... ] TABLE [ ... ] table_name ( [
    { column_name data_type ~\colorbox{red}{[ compress\_mode ]}~  ...  ] )
    ~\colorbox{yellow}{[ INHERITS ( parent\_table [, ... ] ) ] }~
    ~\colorbox{yellow}{[ USING method ]}~
    ...
\end{lstlisting}

\begin{mdframed}
\textit{Solution:}

\textit{Improve robustness with dictionaries and relaxed syntax checks.}
\end{mdframed}

The robustness of syntax-based generation itself can be improved by relaxing the internal checks. Take Squirrel as an example, a test case is generated in three passes: first, the original content is parsed and mutated on AST; next, the generated AST is dumped as a SQL query; finally, the query is re-parsed to ensure its validity. Any error in these steps lead to a failed generation. We take a robustness-oriented design: to enrich the AST corpus, we collect a part of valid AST from those interesting test cases, even if they failed to pass syntax checks as a whole statement.

The robustness can be additionally improved with a customized dictionary. For a specific DBMS, customizing a AST parser to cover all semantic features is labor-consuming and fragile to trivial errors, while customizing a dictionary to effectively guide fuzzing is viable -- it only needs to collect all keywords of the SQL dialect. Although mutation-based methods will generate a number of syntactically-incorrect test cases, its robustness is still attractive in industrial environments.

\subsection{Unreproducible Bugs}

DBMS is a system with rich states. For example, the data can be placed in different shards in a distributed configuration. The cache for query can be ready or be invalidated by other modifications. The statistics of the table could be altered, leading to different query plans. When the system is in a different state, even for executing the same SQL statement, the final execution paths may be vastly different.

Because fuzzing generates a large number of random inputs, the state of the DBMS is significantly altered after fuzzing. Although Squirrel and other testing tools have recorded the specific SQL statement input causing the crash, it is very difficult to reproduce anomalies through the crash-triggering inputs. Sometimes, re-running the same sequence of inputs still cannot reproduce the bug due to the randomness from the operating system.

One useful tool is core dump, which is a file storing a program's state of the working memory, processor state, memory management information, stack pointer, etc. A core dump is generated by the operating system when a program has crashed or terminated abnormally. However, generating a core dump of an enterprise-level DBMS is time-consuming ($\sim$30s) and space-taking ($\sim$500MB) in our cases: enterprise DBMS is optimized for servers with rich compute resources, thus the size of state stored in memory is considerable. Besides, if a bug is discovered, it will be repeatedly triggered in a short period due to the nature of fuzzing and a large number core dump files will be written to the file system. The creation of duplicated core dumps burdens both the server and the developer.
% Moreover, the bug behind many new anomalies is the same. For example, SQLancer produces more than 1 million reports after running for 12 hours. It would be a waste of resources if developers reproduce all the anomalies one by one.

\begin{figure}[htbp]
    \center
    \includegraphics[width=0.8\linewidth]{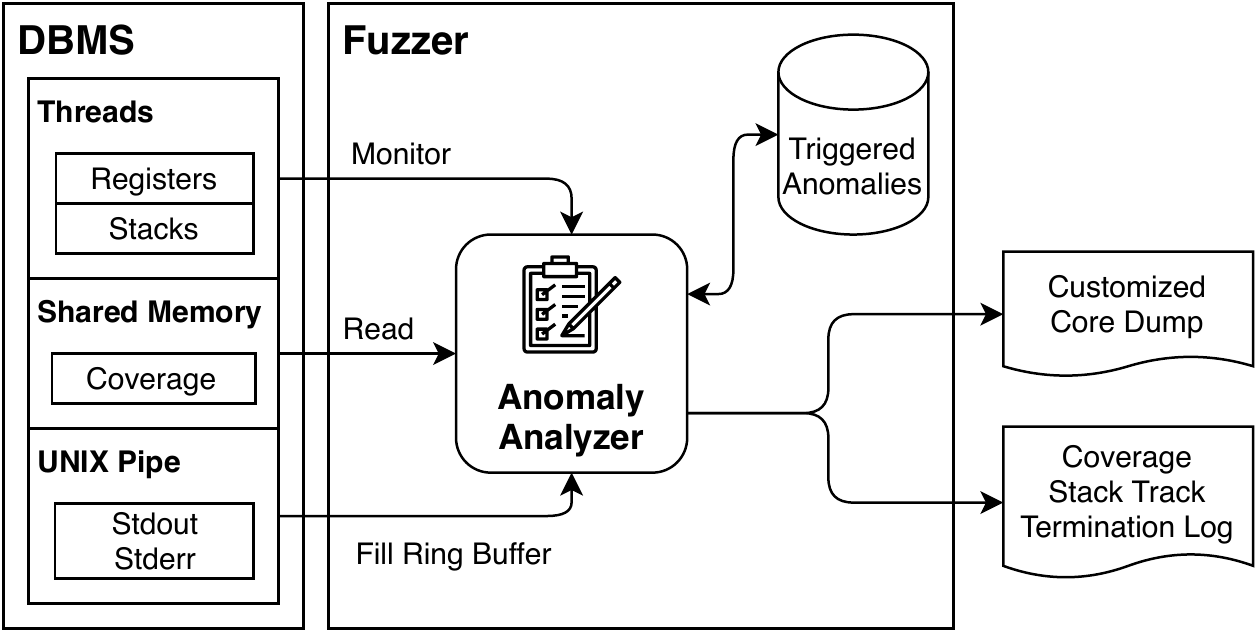}
    \caption{Interaction between the anomaly analyzer and DBMS.}
    \label{fig:monitor}
\end{figure}

\begin{figure}[htbp]
\begin{mdframed}
\textit{Solution:}

\textit{Investigate the root cause of anomalies with on-line triage with feedback-driven deduplication.}
\end{mdframed}
\end{figure}

We design a new scheme to monitor, duplicate, and analyze anomalies, as shown in Fig.~\ref{fig:monitor}.
When a DBMS thread is spawned, the analyzer will trace all its threads by listening for signals. Additionally, it also keeps a ring buffer to store the last output of standard outputs. Once an anomaly is caught, the monitor will collect the rich context where the anomaly happens, namely the coverage feedback, stack trace, and termination log. 
Then the analyzer will remove duplicated anomalies based on coverage and stack trace by querying the set of already-triggered ones. For each unique bug, the analyzer saves the corresponding input, stack trace, original coverage, termination log and a customized core dump.
In this way, the root cause of anomalies can be analyzed online preliminary. For most bugs, developers can fix just with the report, sometimes with the help of core dump. The dependence on reproduction is greatly reduced.

\section{Analysis of Results and Bugs}
Based on the above solutions, we implement a DBMS fuzzer namely \TOOLNAME{}. Our tool mainly consists of three parts: a precise coverage collector, a robust query generator, and an online bug analyzer. 
The precise coverage collector includes two mechanisms, i.e. inter-binary coverage linkage and bijective block mapping. It is implemented on LLVM with 2,092 lines of code in C++. 
The robust query generator is on the top of Rui et al. \cite{SQUIRREL}. We ported its primitive function of parsing and generation (\texttt{parser} and \texttt{src}) to Rust, with 736 lines of C++ as glue. We wrote 2,323 lines of Rust for the overall algorithm.
The online bug analyzer leverages \texttt{ptrace} to monitor behaviors of DBMS's processes. Once an anomaly is triggered, the analyzer will collect the program context ( i.e. on-site stack trace and coverage), check whether the anomaly is newly-triggered based on the context, and disable core dump generation if the anomaly is duplicated. We implement the analyzer with 1,790 lines of code in Rust.

\subsection{Analysis of Results}
Aiming at improving the correctness of our production database GaussDB, we explored various mainstream fuzzing techniques. As Table~\ref{tab:chosen-fuzzer} shows, besides \TOOLNAME{}, we used SQLsmith and SQLancer for grammar-based techniques with and without semantic checks. We also used the academic mutation-based fuzzer Squirrel. We further included several projects for comparison use: PostgreSQL, the original DBMS version where GaussDB forked from; Comdb2, another enterprise-level DBMS from Bloomberg LP.

We perform our evaluation on a 64-bit machine with 40 cores (Intel(R) Xeon(R) Gold 6148 CPU @ 2.40GHz), 128 GiB of RAM, 2 * 7,200 rpm hard disk in RAID0, and Linux 5.5.13.
All DBMSs were compiled with \texttt{-O2} flags. We enable sanitizers for enhanced bug-detection ability on fuzzers. The initial seeds were collected from the built-in integration tests. 
We conducted each experiment for 12 hours. We collected the seeds generated by those fuzzers, and dry-ran the seeds through our instrumentation to compute their coverage.
Note that due to Comdb2 heavily uses Lua to extend its SQLite-based SQL dialect, we failed to adapt Squirrel and SQLsmith to it within a reasonable time. 
In addition, we do many engineering efforts to adapt Squirrel to GaussDB. But after Squirrel covered 229 blocks, the fuzzer crashed due to its implementation issues. As a result, we skipped the evaluation of SQLsmith and Squirrel on Comdb2 and Squirrel on GaussDB.

\begin{table}[htbp]
    \caption{\label{tab:coverage} Coverage of DBMS for each Fuzzer}
    \centering
    \small
    % \resizebox{1\textwidth}{7mm} {
    \begin{tabular}{l|rrrr}
        \toprule
        DBMS      & SQLsmith & SQLancer & Squirrel & \TOOLNAME{} \\ 
        \hline
        GaussDB    & 50,172 &  2,513 &    N/A  & 69,432 \\ 
        PostgreSQL & 42,563 & 39,913 & 16,954  & 87,739 \\
        Comdb2     & N/A    &  2,773 &    N/A  & 18,941 \\
        \bottomrule
    \end{tabular}
    % }
\end{table}

Table~\ref{tab:coverage} presents the number of covered basic blocks for each fuzzer. 
The second and third columns present the results for generation-based fuzzers, and the last two columns present the results for mutation-based fuzzer. 
From the table, we can see that: 

\textbf{Simple blackbox fuzzers have good robustness in industrial practice.} For example, the blackbox fuzzer SQLsmith outperformed the coverage-guided fuzzer Squirrel by 151.05\%. It is because Squirrel is an academic prototype which lacks feedback from the industry while SQLsmith is designed for testing PostgreSQL from the beginning. 

\textbf{Complex blackbox methods have limitations in industrial practice.} For example, SQLancer covered more basic blocks than Squirrel on PostgreSQL but under-performed SQLsmith on GaussDB and PostgreSQL. The reason is that SQLancer is specialized in logic bug hunting and only uses a part of SQL features to construct its oracle violation. Its strength is discovering more types of bugs instead of coverage improvement. SQLancer also uses advanced test oracles to detect bugs, thus the \texttt{CREATE TABLE} statements generated by it have many fine-tuned parameters. It performs well on its supported DBMSs such as PostgreSQL and SQLite. However, when adapting to PostgreSQL-based GaussDB and SQLite-based Comdb2, the trivial differences in syntax prevented it from function normally. 

\textbf{\TOOLNAME{} outperformed other fuzzers in terms of coverage and adaptation range.} On GaussDB, it covers 0.38x, 26.63x more basic blocks than SQLsmith and SQLancer, respectively; On PostgreSQL, it covers 1.06x, 1.20x, 4.29x more basic blocks than SQLsmith, SQLancer, and Squirrel, respectively; On Comdb2, it covers 5.83x than SQLancer.

\begin{figure}[htbp]
    \center
    \includegraphics[width=\linewidth]{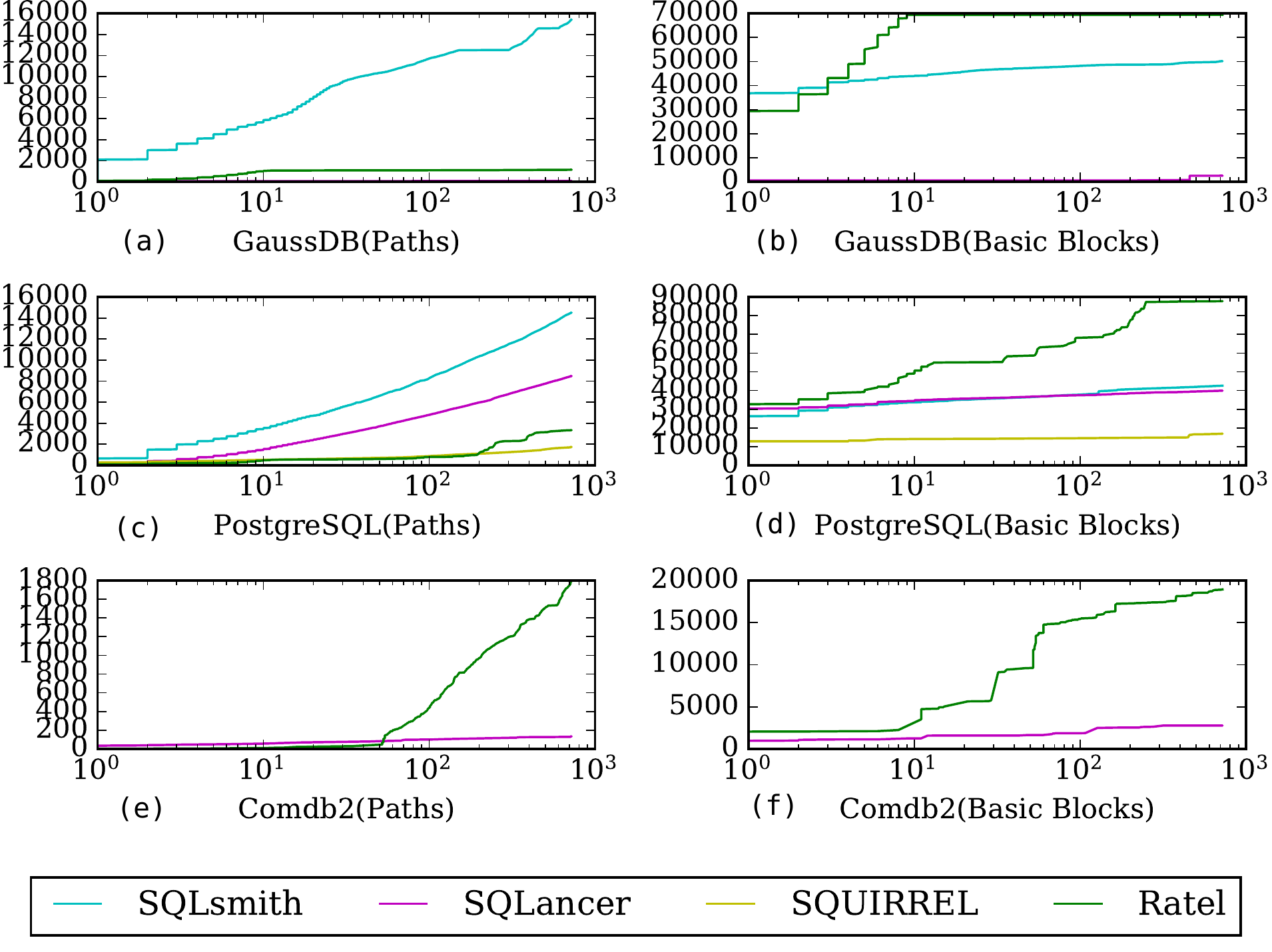}
    \caption{Path and basic block growth for each fuzzer over the 12-hour trials.}
    \label{fig:trends}
\end{figure}

Figure~\ref{fig:trends} demonstrates the number of basic blocks and paths growth for each fuzzer over the 12-hour trails in the first and the second column, respectively. 
The results illustrate that \TOOLNAME{} not only achieves high coverage, but also has advantages in speed.
In specific, in Figure~\ref{fig:trends} (a) and Figure~\ref{fig:trends} (b), \TOOLNAME{} finds less basic blocks and paths than SQLsmith in the beginning. But with the coverage guided, \TOOLNAME{} quickly catches up with and surpasses it. In Figure~\ref{fig:trends} (c) and Figure~\ref{fig:trends} (d), \TOOLNAME{} covers more basic blocks and paths in PostgreSQL than other fuzzers in the beginning, and it maintains lead to the end. The results on Comdb2 in Figure~\ref{fig:trends} (e) and Figure~\ref{fig:trends} (f) also have similar trends.

The performance promotion of \TOOLNAME{} mainly dues to high-quality coverage guidance. With the coverage guidance, Squirrel and \TOOLNAME{} find more basic blocks significantly. For example, both the curves of Squirrel and \TOOLNAME{}  in Figure~\ref{fig:trends} are always above the other black box fuzzers. 
Compared to Squirrel, \TOOLNAME{} still performs better because of the more precise feedback. Squirrel employs the feedback mechanism in AFL. Because DBMSs always have a large code scale, Squirrel extends the bitmap from 64 KB to 256KB. However, the AFL's hash collision issue is still serious and unavoidable. For example, PostgreSQL has 354K basic blocks. 
According to CollAFL\cite{collafl}, a smaller project libtorrent (which has 164K basic blocks) still has about a 40\% collision rate with 256KB. In addition, expanding bitmap size will slow down the execution speed. For example, the speed of libtorrent slows down to around 80\%. For DBMSs like PostgreSQL, the collision will be more significant, and the overhead will have a more serious impact.

In contrast, \TOOLNAME{} implements conflict-free cross-binary feedback. It eliminates the collision by assigning unique counters for each basic block. To enrich block-based coverage, \TOOLNAME{} analyzes critical edges, whose source has multiple successors and destination has multiple predecessors and represents them with dummy basic blocks. Therefore, the improved coverage brings more opportunities for \TOOLNAME{} to find more vulnerabilities.

 \subsection{Analysis of Bugs}
In our practice of fuzzing DBMSs, \TOOLNAME{} found 32, 42, and 5 previously-unknown bugs in GaussDB, PostgreSQL, and Comdb2. In contrast, SQLancer, SQLsmith, and Squirrel did not find them. Out of these bugs, we take two interesting cases from these DBMSs. For both cases, we illustrate how the challenges in adapting coverage-feedback fuzzing to DBMSs impedes other fuzzer find these bugs. And more importantly, we also analyze how the solutions help \TOOLNAME{} to find them.

\paragraph{Null Pointer Access in PostgreSQL}
 
\begin{figure}
\begin{lstlisting}[language={SQL},frame=single, rulesepcolor=\color{red!20!green!20!blue!20},xrightmargin=-0.5em, aboveskip=1em,caption={The SQL statements triggering null pointer access.},label=bug-pg]
-- Terminate the corresponding backend process.
SELECT ~\colorbox{yellow}{pg\_cancel\_backend(pg\_backend\_pid())}~ 
-- Now the connection is closed by the server.
\c non_existing_db -- Tries to reuse the closed connection.
~\colorbox{color_error}{SELECT ...}~ -- CRASH WITH NULL POINTER!
\end{lstlisting}
\end{figure}

% \paragraph{Case study bugxxx, which occurs when the client of PostgreSQL reused the old connection to server}
Listing \ref{bug-pg} shows the SQL statements which could trigger the bug. This bug involves multiple interactions between the server and the client, which could be triggered with three statements.
In specific, first the SQL statement \texttt{SELECT pg\_cancel\_backend(pg\_backend\_PID ())} is executed, and the \texttt{backend} process of the server will shut down and close the connection.
Next, the client tries to switch to another database by attempting to reuse the old connection, but the connection is closed and set to NULL pointer. Next, executing any SQL statements tries to access the NULL connection, crashing the client as a result.

Both SQLancer and SQLsmith did not find this bug. They use blackbox methods, which build syntactically correct SQLs based on ASTs. However, queries like \texttt{pg\_cancel\_backend} are DBMS specific extension functions.
% which are different across DBMSs. Different DBMSs often have their own specific extensions to rich their features. To create a model which covers all these features is difficult and requires a lot of manual efforts.
Due to the incomplete syntax modeling, SQLancer and SQLsmith can never generate such queries.
Squirrel uses syntax-based mutation to generate SQL statements and enforces strict language-validity: only the inputs which pass its AST parser could be mutated, and the generated query is re-parsed to ensure the validity. However, the incomplete modeling of the syntax prevented it from mutating the input at all. Consequently, neither can Squirrel trigger this bug.

Different from these fuzzers, \TOOLNAME{} addresses the challenge of fragile input generation by not only adapting PostgreSQL with a customized dictionary, but also preserving the syntax-based mutation with relaxed syntax checks. Because it only needs to collect all keywords of the SQL dialect, the workload is decreased while the performance improves. More importantly, \TOOLNAME{} is robustness-oriented. Even if some interesting generated cases failed to pass syntax checks, we still collect them for further mutation. As a result, the bug was finally found by \TOOLNAME{}.

\paragraph{Use-After-Scope in Comdb2}
This bug could be triggered only when Comdb2 in a specific state. As Fig.~\ref{fig:comdb2} shows, this bug resides in networking code, but can only be triggered within the context of schema change:

\begin{figure}[htbp]
    \center
    \includegraphics[width=0.8\linewidth]{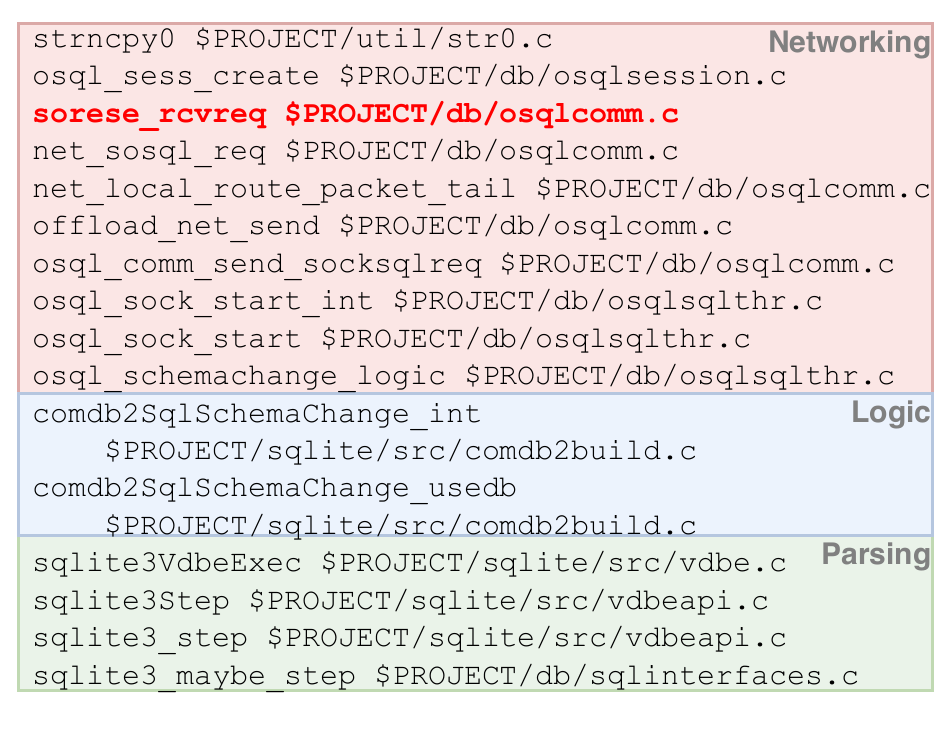}
    \caption{Stack trace captured by \TOOLNAME{} when the use-after-scope happened. The shortened version is presented here for brevity.}
    \label{fig:comdb2}
\end{figure}

\begin{enumerate}
    \item An \texttt{ALTER} statement is successfully parsed and converted to the bytecode of Virtual Database Engine.
    \item The semantic check of schema change is passed, and the logic for coordinating the schema change is executed.
    \item The schema change is decomposed to a series of operations, including a \texttt{uuid} networking request.
\end{enumerate}

Solely with the stack trace, developers can analyze and fix the bug. Listing \ref{bug-comdb2} shows the code snippets of the bug in function \texttt{sorese\_rcvreq} in Step 3.
First, when the node sends a \texttt{uuid-request}, \texttt{tzname} is assigned with the value of a local variable \texttt{ureq}. Second, the \texttt{tzname} is used to create \texttt{sess} in function \texttt{osql\_sess\_create}. However, the local variable \texttt{ureq} is inaccessible at this scope. In other words, \texttt{tzname} points to an invalid stack address. As a result, Comdb2 encounters a stack use after scope error and crashes.

\begin{figure}[htbp]
\begin{lstlisting}[language=C++, caption={The details of stack-use-after-scope on address},label=bug-comdb2] 
int sorese_rcvreq(...) {
    char *tzname;
    /* grab the request */
     ~\colorbox{color_myself}{\textsf{The networking code which handles request type of UUID.}}~
    if (osql_nettype_is_uuid(nettype)) {
        ~\colorbox{color_myself}{Note that ureq is a local variable.}~
        osql_uuid_req_t ~\colorbox{color_error}{ureq;}~
        ~\colorbox{color_error}{tzname}~ = ureq.tzname;
       ... 
    } else {
       ...
    }
    /* create the request */
    ~\colorbox{color_myself}{\texttt{ureq} \textsf{does not live long enough.}}~
    ~\colorbox{color_myself}{\textsf{Now \texttt{tzname} points to a invalid address.}}~
    sess = osql_sess_create(..., ~\colorbox{red}{tzname}~,...);
    .....
}

osql_sess_t *osql_sess_create(){
    if (tzname)
      ~\colorbox{color_myself}{\textsf{Crash due to invalid memory access.}}~
      strncpy0(sess->tzname, ~\colorbox{red}{tzname}~, sizeof(sess->tzname));
}
\end{lstlisting}
\end{figure}

The bug is hidden in deep logic, which is triggered only when the specific basic blocks are covered and Comdb2 reaches a unique state after querying a particular \texttt{ALTER} statement. As shown in Fig.~\ref{fig:comdb2}, the call stack of the bug is rather deep. In specific, it contains three levels, namely parsing, logic, and networking. Covering the basic blocks on the top of the stack is hard. For blackbox fuzzers like SQLsmith and SQLancer, they have a low possibility of generating SQL statements to cover these states without feedback. Although Squirrel also guides fuzzing with coverage, the spare coverage and its strict syntax checks lower the probability to trigger the bug.
In contrast, \TOOLNAME{} addresses the challenges of imprecise coverage collection with inter-binary linkage and bijective block mapping. With the robustness-oriented syntax-based mutation, it expanded coverage and finally triggered this bug.

More importantly, this bug requires Comdb2 in a particular state: just passing the syntax checks is not enough, the table must be present and its schema must be valid for the \texttt{ALTER} statement. Therefore, simply re-executing the crash-triggering input cannot meet the requirement of the table's state. For other fuzzers, even if recording all the executed SQL statements, it still takes great human efforts to reproduce and pinpoint the root cause.
Unlike other fuzzers, \TOOLNAME{} addresses the challenges of unreproducible bugs with on-line triage. It monitors and catches the anomaly, saves the corresponding input, termination log, and a customized core dump. With detailed information, maintainers can analyze the cause of the bug without requiring to reproduce the anomaly.

\section{Lessons Learned}
In this section, we introduce some lessons learned on DBMS fuzzing deployment.

\paragraph{Importance of Initial Seeds}
The richness of initial seeds plays a vital role in bug discovery. Fuzzer will not waste time to explore shallow and rapidly trigger deep logic in DBMSs if initial seeds cover more functionalities. Nonetheless, manually writing SQL statements as initial seeds is labor-consuming. In our practice, we found the integration tests include a variety of SQL statements, even for corner cases. We collected seeds from these tests for better performance.

\paragraph{Lack of Supports for Parallel Mode}
The industry deploys thousands of fuzzing instances. To optimize resource usage, DBMS fuzzing can be performed by spawning multiple fuzzing clients to connect to a server and pass data.
Blackbox DBMS fuzzers support parallel mode well while Squirrel cannot follow this pattern. It is because 
Squirrel should maintain program states when collecting coverage. If other clients send queries during the interval, the collected coverage is imprecise. In our practice, we can only set up the equivelant number of servers as the number of clients for Squirrel, which is ineffective. We believe it is a valuable research topic and demand to be tackled.

\paragraph{Low Efficiency in Dropping Databases}
Dropping database decreases the noise for stateful DBMSs, which assists bug deduplication and helps fuzzer performs more stable. However, frequent dropping will ignore the bugs which only can be triggered in some states and bring a lot of overhead. For example, Squirrel and SQLancer drop databases after each query. We occasionally drop databases to explore more states and lower the overhead. Instead, when bugs are triggered, we investigate the root cause of bugs with on-line triage.

\paragraph{Bugs Hidden in Heterogeneous Platforms}
In practice, we find bugs hidden in heterogeneous platforms. For example, GaussDB is mainly deployed in Huawei Kunpeng powered ARM servers. We find that some bugs like misaligned pointer accesses have minor performance impacts on x86 platforms, but can cause hardware exceptions on ARM. Undefined behaviors might seem trivial on some certain platforms, they can still lead to major bugs in other platforms.

\paragraph{Difficulty in Confirming Bugs}
We found many bugs after running fuzz testing. However, we encounter resistance in confirming them. Although the bugs found by fuzzing are unlikely to be false-positives, for complicated stateful DBMSs, many of these bugs are hard to reproduce. As a result, some of the maintainers tend to refuse to confirm them. To lower the difficulty of confirming bugs, the tester needs to supply more detailed information about them. In our practice, besides recording input seeds that cause anomaly like other fuzzers, we also save the core dump file as well as the context (e.g. stack trace and termination log). This gives the maintainers more evidence to analyze and confirm the bug. Nonetheless, raising the security awareness of the maintainers is still a long way to go.

\section{Conclusion}
In this paper, we present the practice of adapting coverage-guided fuzzing to enterprise-level DBMSs from Huawei and Bloomberg LP, which differs greatly from fuzzing library-level DBMSs like SQLite. The difficulty is due to extra complexity and distributed nature. We encountered three main challenges, i.e. imprecise coverage collection, fragile input generation, and unreproducible bugs. We discuss the solutions for each challenge and propose \TOOLNAME{}, a coverage-guided fuzzer for enterprise-level DBMSs. It uses an industry-oriented design, which improves the feedback precision, enhances SQL generation, and performs on-line bug analysis.
We used \TOOLNAME{} to test GaussDB, Comdb2, and PostgreSQL, and 32, 42, and 5 unknown bugs are discovered respectively. Hoping for a better adaptation of fuzzing in enterprise-level DBMSs, we further summarize them into valuable lessons.

\section*{Acknowledgement}
This research is sponsored in part by the NSFC Program (No. 62022046, U1911401, 61802223), National Key Research and Development Project (Grant No. 2019YFB1706200) and Ali-Tsinghua Database Testing Research Project (NO. 20212000070).

\bibliographystyle{IEEEtran}
\bibliography{industry}

\end{document}

%% file: mine.tex
\usepackage{comment}
\usepackage{xcolor}
\usepackage{enumitem}
\usepackage{adjustbox}
\usepackage{tabularx}

\usepackage{listings}
\definecolor{color_comment}{rgb}{0.0, 0.5, 0.0}
\definecolor{color_keyword}{rgb}{0.5, 0.0, 0.0}
\definecolor{color_myself}{rgb}{0.9, 0.9, 0.9}
\definecolor{color_error}{rgb}{1, 0.2, 0.2}

%% file: industry.bbl
% Generated by IEEEtran.bst, version: 1.14 (2015/08/26)
\begin{thebibliography}{10}
\providecommand{\url}[1]{#1}
\csname url@samestyle\endcsname
\providecommand{\newblock}{\relax}
\providecommand{\bibinfo}[2]{#2}
\providecommand{\BIBentrySTDinterwordspacing}{\spaceskip=0pt\relax}
\providecommand{\BIBentryALTinterwordstretchfactor}{4}
\providecommand{\BIBentryALTinterwordspacing}{\spaceskip=\fontdimen2\font plus
\BIBentryALTinterwordstretchfactor\fontdimen3\font minus
  \fontdimen4\font\relax}
\providecommand{\BIBforeignlanguage}[2]{{%
\expandafter\ifx\csname l@#1\endcsname\relax
\typeout{** WARNING: IEEEtran.bst: No hyphenation pattern has been}%
\typeout{** loaded for the language `#1'. Using the pattern for}%
\typeout{** the default language instead.}%
\else
\language=\csname l@#1\endcsname
\fi
#2}}
\providecommand{\BIBdecl}{\relax}
\BIBdecl

\bibitem{SQLsmith}
\BIBentryALTinterwordspacing
A.~Seltenreich, B.~Tang, and S.~Mullender, ``Sqlsmith: a random sql query
  generator,'' 2018. [Online]. Available:
  \url{https://github.com/anse1/sqlsmith}
\BIBentrySTDinterwordspacing

\bibitem{DBLP:conf/vldb/Slutz98}
D.~R. Slutz, ``Massive stochastic testing of {SQL},'' in \emph{VLDB'98,
  Proceedings of 24rd International Conference on Very Large Data Bases, New
  York, {USA}}.\hskip 1em plus 0.5em minus 0.4em\relax Morgan Kaufmann, pp.
  618--622.

\bibitem{SQLancer}
\BIBentryALTinterwordspacing
M.~Rigger, ``Sqlancer: detecting logic bugs in dbms,'' 2020. [Online].
  Available: \url{https://github.com/sqlancer/sqlancer}
\BIBentrySTDinterwordspacing

\bibitem{NoREC}
M.~Rigger and Z.~Su, ``{Detecting Optimization Bugs in Database Engines via
  Non-Optimizing Reference Engine Construction},'' in \emph{Proceedings of the
  2020 28th ACM Joint Meeting on European Software Engineering Conference and
  Symposium on the Foundations of Software Engineering}, ser. ESEC/FSE 2020,
  2020.

\bibitem{PQS}
\BIBentryALTinterwordspacing
``Testing database engines via pivoted query synthesis,'' in \emph{14th
  {USENIX} Symposium on Operating Systems Design and Implementation ({OSDI}
  20)}.\hskip 1em plus 0.5em minus 0.4em\relax Banff, Alberta: {USENIX}
  Association, Nov. 2020. [Online]. Available:
  \url{https://www.usenix.org/conference/osdi20/presentation/rigger}
\BIBentrySTDinterwordspacing

\bibitem{PAFL}
J.~Liang, Y.~Jiang, Y.~Chen, M.~Wang, C.~Zhou, and J.~Sun, ``{PAFL:} extend
  fuzzing optimizations of single mode to industrial parallel mode,'' in
  \emph{Proceedings of the 2018 {ACM} Joint Meeting on European Software
  Engineering Conference and Symposium on the Foundations of Software
  Engineering, {ESEC/SIGSOFT} {FSE}, USA, November 04-09, 2018}, 2018, pp.
  809--814.

\bibitem{zeror}
C.~Zhou, M.~Wang, J.~Liang, Z.~Liu, and Y.~Jiang, ``Zeror: Speed up fuzzing
  with coverage-sensitive tracing and scheduling,'' in \emph{2020 35th IEEE/ACM
  International Conference on Automated Software Engineering (ASE)}.\hskip 1em
  plus 0.5em minus 0.4em\relax IEEE, 2020, pp. 858--870.

\bibitem{SQUIRREL}
R.~Zhong, Y.~Chen, H.~Hu, H.~Zhang, W.~Lee, and D.~Wu, ``Squirrel: Testing
  database management systems with language validity and coverage feedback,''
  in \emph{The ACM Conference on Computer and Communications Security (CCS),
  2020}, 2020.

\bibitem{SQLite-Bugs}
``{SQLite: All Tickets},'' \url{https://www.sqlite.org/src/rptview?rn=1}, 2020.

\bibitem{PostgreSQL-Bugs}
``{PostgreSQL: PostgreSQL Mailing Lists: pgsql-bugs},''
  \url{https://www.postgresql.org/list/pgsql-bugs/}, 2020.

\bibitem{LLVM}
C.~Lattner and V.~Adve, ``Llvm: A compilation framework for lifelong program
  analysis \& transformation,'' in \emph{International Symposium on Code
  Generation and Optimization, 2004. CGO 2004.}\hskip 1em plus 0.5em minus
  0.4em\relax IEEE, 2004, pp. 75--86.

\bibitem{gaussdb-doc}
``{Gaussdb Documentation},''
  \url{https://opengauss.org/en/docs/1.0.1/docs/Releasenotes/Terms-of-Use.html},
  2020.

\bibitem{DBLP:journals/pvldb/AvniAAABGGLLLMM20}
H.~Avni, A.~Aliev, O.~Amor, A.~Avitzur, I.~Bronshtein, E.~Ginot, S.~Goikhman,
  E.~Levy, I.~Levy, F.~Lu, L.~Mishali, Y.~Mo, N.~Pachter, D.~Sivov,
  V.~Veeraraghavan, V.~Vexler, L.~Wang, and P.~Wang, ``Industrial strength
  {OLTP} using main memory and many cores,'' \emph{Proc. {VLDB} Endow.},
  vol.~13, no.~12, pp. 3099--3111, 2020.

\bibitem{TLP}
M.~Rigger and Z.~Su, ``Finding bugs in database systems via query
  partitioning,'' \emph{Proc. ACM Program. Lang.}, no. OOPSLA, 2020.

\bibitem{jung:apollo}
J.~Jung, H.~Hu, J.~Arulraj, T.~Kim, and W.~Kang, ``{APOLLO: Automatic Detection
  and Diagnosis of Performance Regressions in Database Systems (to appear)},''
  in \emph{Proceedings of the 46th International Conference on Very Large Data
  Bases (VLDB)}, Tokyo, Japan, Aug. 2020.

\bibitem{SQLsmith_score_list}
\BIBentryALTinterwordspacing
A.~Seltenreich, B.~Tang, and S.~Mullender, ``Sqlsmith: Score list,'' 2020.
  [Online]. Available: \url{https://github.com/anse1/sqlsmith/wiki#score-list}
\BIBentrySTDinterwordspacing

\bibitem{sqlancer_score_list}
\BIBentryALTinterwordspacing
M.~Rigger, ``Sqlancer: Bugs found in database management systems,'' 2020.
  [Online]. Available: \url{https://www.manuelrigger.at/dbms-bugs}
\BIBentrySTDinterwordspacing

\bibitem{afl}
\BIBentryALTinterwordspacing
lcamtuf, ``American fuzzy lop (afl),'' 2017. [Online]. Available:
  \url{http://lcamtuf.coredump.cx/afl/}
\BIBentrySTDinterwordspacing

\bibitem{libFuzzer}
\BIBentryALTinterwordspacing
libfuzzer@googlegroups.com, ``libfuzzer – a library for coverage-guided fuzz
  testing.'' 2020. [Online]. Available:
  \url{https://llvm.org/docs/LibFuzzer.html}
\BIBentrySTDinterwordspacing

\bibitem{honggfuzz}
\BIBentryALTinterwordspacing
honggfuzz@googlegroups.com, ``hongfuzz - security oriented fuzzer with powerful
  analysis options.'' 2020. [Online]. Available: \url{http://honggfuzz.com}
\BIBentrySTDinterwordspacing

\bibitem{DBLP:conf/icse/SAFL}
M.~Wang, J.~Liang, Y.~Chen, Y.~Jiang, X.~Jiao, H.~Liu, X.~Zhao, and J.~Sun,
  ``{SAFL:} increasing and accelerating testing coverage with symbolic
  execution and guided fuzzing,'' in \emph{Proceedings of the 40th
  International Conference on Software Engineering: Companion Proceeedings,
  {ICSE} 2018, Gothenburg, Sweden}.\hskip 1em plus 0.5em minus 0.4em\relax
  {ACM}, 2018, pp. 61--64.

\bibitem{DBLP:conf/ndss/Driller}
N.~Stephens, J.~Grosen, C.~Salls, A.~Dutcher, R.~Wang, J.~Corbetta,
  Y.~Shoshitaishvili, C.~Kruegel, and G.~Vigna, ``Driller: Augmenting fuzzing
  through selective symbolic execution,'' in \emph{23rd Annual Network and
  Distributed System Security Symposium, {NDSS} 2016, San Diego, California,
  USA, February 21-24, 2016}.\hskip 1em plus 0.5em minus 0.4em\relax The
  Internet Society, 2016.

\bibitem{DBLP:conf/kbse/FairFuzz}
\BIBentryALTinterwordspacing
C.~Lemieux and K.~Sen, ``Fairfuzz: a targeted mutation strategy for increasing
  greybox fuzz testing coverage,'' in \emph{Proceedings of the 33rd {ACM/IEEE}
  International Conference on Automated Software Engineering, {ASE} 2018,
  Montpellier, France, September 3-7, 2018}, 2018, pp. 475--485. [Online].
  Available: \url{https://doi.org/10.1145/3238147.3238176}
\BIBentrySTDinterwordspacing

\bibitem{InteFuzz}
J.~Liang, Y.~Chen, M.~Wang, Y.~Jiang, Z.~Yang, C.~Sun, X.~Jiao, and J.~Sun,
  ``Engineering a better fuzzer with synergically integrated optimizations,''
  in \emph{2019 IEEE 30th International Symposium on Software Reliability
  Engineering (ISSRE)}.\hskip 1em plus 0.5em minus 0.4em\relax IEEE, 2019, pp.
  82--92.

\bibitem{iso_standard}
``{STANDARDS BY ISO/IEC JTC 1/SC 32 Data management and interchange},''
  \url{https://www.iso.org/committee/45342/x/catalogue/p/1/u/0/w/0/d/0}, 2020.

\bibitem{collafl}
\BIBentryALTinterwordspacing
S.~Gan, C.~Zhang, X.~Qin, X.~Tu, K.~Li, Z.~Pei, and Z.~Chen, ``Collafl: Path
  sensitive fuzzing,'' in \emph{2018 {IEEE} Symposium on Security and Privacy,
  {SP} 2018, Proceedings, 21-23 May 2018, San Francisco, California,
  {USA}}.\hskip 1em plus 0.5em minus 0.4em\relax {IEEE} Computer Society, 2018,
  pp. 679--696. [Online]. Available:
  \url{https://doi.org/10.1109/SP.2018.00040}
\BIBentrySTDinterwordspacing

\end{thebibliography}
